\documentclass[aps,prl,twocolumn,superscriptaddress,showpacs,showkeys]{revtex4}

\usepackage{graphicx}

\begin{document}


\title{Mesoscopic Magnetic States in Metallic Alloys with Strong Electronic Correlations: A Percolative Scenario for CeNi$_{1-x}$Cu$_{x}$ }



\author{N. Marcano}
\affiliation{Dpto. CITIMAC, Universidad de Cantabria, 39005 Santander, Spain}
\affiliation{Cavendish Laboratory, University of Cambridge, Cambridge CB3 OHE, United Kingdom}
\author{J.C. G\'omez Sal}
\author{J.I. Espeso}
\affiliation{Dpto. CITIMAC, Universidad de Cantabria, 39005 Santander, Spain}
\author{J.M. De Teresa}
\author{P.A. Algarabel}
\affiliation{ICMA, CSIC - Universidad de Zaragoza, 50009 Zaragoza, Spain}
\author{C. Paulsen}
\affiliation{Centre de Recherche sur les Tr\`es Basses Temp\'eratures, CNRS, 38042 Grenoble, France }
\author{J.R. Iglesias}
\affiliation{Instituto de F\'{\i}sica, Universidade Federal do Rio Grande do Sul, 91501-970 Porto Alegre, Brazil}



\date{\today}

\begin{abstract}
We present evidence for the existence of magnetic clusters  of approximately  20 \AA \ in the strongly correlated alloy system CeNi$_{1-x}$Cu$_{x}$ (0.7 $\le$ x $\le$ 0.2) based on small angle neutron scattering experiments as well as the  occurrence  of staircase-like hysteresis cycles during very low temperature (100 mK) magnetization measurements. An unusual feature is the observation of long-range ferromagnetic order below the cluster-glass transition without any indication of a sharp transition at a Curie temperature. These observations strongly support a phenomenological model where a percolative process connects the cluster-glass state observed at high temperatures with the long-range ferromagnetic order observed by neutron diffraction experiments at very low temperatures. The model can account for all the puzzling macroscopic and microscopic data previously obtained in this system, providing a new perspective with regard to the magnetic ground state of other alloyed compounds with small magnetic moments or weak ferromagnetism with intrinsic disorder effects. 
\end{abstract}


\maketitle


In order to investigate the physical and, in particular, magnetic properties of materials it is often common and useful to substitute different species of ions into metals and alloys. Ability to tune to the properties of matter in this manner allows one to distinguish between the applicability and limitations of different theoretical models to describe the emergent phenomena.  For example, one of the first experimental demonstration of the Doniach's phase diagram was from the systematic study of Pt substitutions in the CeNi$_{1-x}$Pt$_{x}$ series \cite{Gignoux,Blanco}. Doniach's phenomenology explains the competition between the RKKY interaction and the conduction band hybridization in Cerium and Uranium compounds some of them exhibiting Non-Fermi Liquid (NFL) behaviour in the proximity of a Quantum Critical Point (QCP) \cite{Stewart}.  In some important examples, substitution or doping can improve a desired property, as is the case for materials such as High-T$_{C}$ \cite{Pan} and other unconventional superconductors \cite{Steglich,Yuan} as well colossal magneto-resistance (CMR) compounds \cite{DeTeresa,Burgy}.

It is clear that strong electron correlations are the origin of the properties for all the examples mentioned above and as a consequence,  many of these properties are sensitive to disorder \cite{Millis,Miranda}. One is thus left with the puzzle of how to discern whether the observed novel phenomena or change in property is due intrinsic inhomogeneities or simply a consequence of the poor quality of sample. It is important to note that regardless of the amount of substitution one can not guarantee that the distribution of the dopant to be uniform across the volume of the sample.  It is indeed possible to achieve crystallographic homogeneity in the doped compounds, however one has to be attentive to the implications at the electronic level. The average effect of such substitutions is clearly observed in macroscopic measurements such as X-ray diffraction, magnetization, specific heat etc, but the heterogeneities only become apparent when microscopic techniques, particularly muon spin relaxation ($\mu$SR) and transmission electron microscopy are used. In this sense, even appropriate annealing which brings about perfect homogenisation is unable to prevent intrinsic inhomogeneities from forming in alloyed and doped compounds. A special effort is thus required in the structural, electrical and magnetic characterization by combining macroscopic and microscopic techniques in order to obtain a complete picture of the magnetic behaviour of such complex systems. 

\begin{figure}
\includegraphics[width=7.5cm]{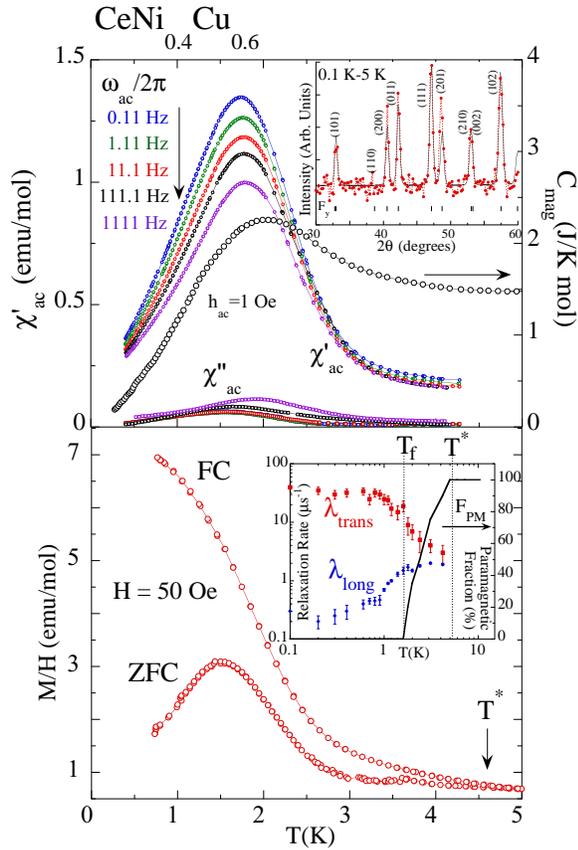}
\caption{Summary of the macroscopic and microscopic measurements for CeNi$_{0.4}$Cu$_{0.6}$: Top: The real ($\chi^{\prime}_{ac}$) and imaginary ($\chi^{\prime\prime}_{ac}$) components of the ac-susceptibility in an applied ac field of 1 Oe at various selected frequencies, and the magnetic contribution to the specific heat $C_{mag}$ in the low temperature regime. The inset shows the magnetic neutron-diffraction pattern (difference between 0.1 and 5 K). Points indicate the experimental data and the line is the calculated pattern. Vertical ticks mark the magnetic reflections for $|\vec{q}|$ = 0. Bottom: Field Cooled and Zero Field Cooled magnetization in a field of 50 Oe. The inset shows the temperature dependence of the $\mu$SR relaxation rates $\lambda_{trans}$ and $\lambda_{long}$, and the paramagnetic volume fraction, $F_{PM}$, in the intermediate regime.}
\end{figure}

Accordingly, a complete study of the CeNi$_{1-x}$Cu$_{x}$ series has been developed in recent years using policrystalline samples \cite{Soldevilla,Espeso,Marcano05,Marcano}. The main objective of this study was the search for a NFL behaviour close to the QCP expected around x = 0.2 (CeCu is antiferromagnetic (AFM) and evolves to ferromagnetism (FM) for x $<$ 0.8, while CeNi is an intermediate valence compound). However, our study revealed puzzling and seemingly contradictory physical results. In particular: (i) Long-range magnetic order has been obtained from neutron diffraction  \cite{Espeso} at low temperatures (AFM for compounds 1 $<$ x $\leq$ 0.7 and FM for compounds with 0.6 $\leq$ x $\leq$ 0.3). In addition, a cluster-glass state was determined in the FM compositions by ac-susceptibility ($\chi_{ac}$) at higher temperatures. The freezing temperature, $T_{f}$, coincides with the maxima observed in specific heat measurements \cite{Marcano05}; (ii) No indication of any Curie temperature, $T_{C}$, associated to the onset of long-range ferromagnetic order was observed by macroscopic measurements ($\chi_{ac}$, $C_{P}$). $\mu$SR results indicate the formation of dynamic spin clusters below a characteristic temperature $T^{*}$, above $T_{f}$  \cite{Marcano}; (iii) considering the whole analysis of the results, the existence of a QCP was not required to describe the evolution of the magnetism along the series, although the low temperature $C_{P}$ for x = 0.2 follows a NFL-like temperature dependence. 
In this letter we present new Small Angle Neutron Scattering (SANS) and very low temperature (100 mK) magnetization data that lead us to propose a new phenomenological model that encompasses all the previous macroscopic and microscopic measurements obtained on this system. This picture is based on a cluster-glass state percolating into long-range ferromagnetic order at low temperature.

\begin{figure}
\includegraphics[width=8cm]{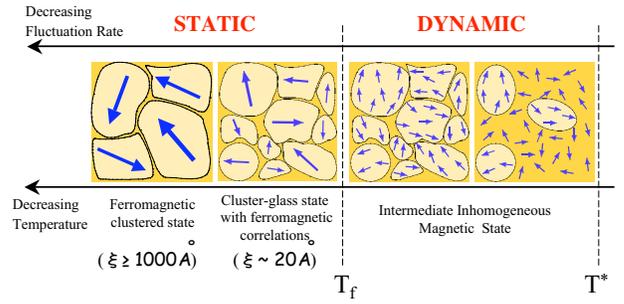}
\caption{Schematic illustration of the magnetic state according to the proposed magnetic cluster model in different temperature regions. $T_{f}$ denotes the cluster-glass freezing temperature and $T^{*}$ the establishment of the intermediate magnetic inhomogeneous state (see text); the latter is determined by $\mu$SR ZF-measurements, and the former from the cusp in ac-susceptibility. Note that these two temperatures evolve with the composition along the series. $\xi$ corresponds to the magnetic correlation length.}
\end{figure}

Figure 1 displays the experimental evidence of the above mentioned behaviour for one characteristic compound (x = 0.6) as a representative sample presenting long-range ferromagnetic order at a very low temperature (see inset of Figure 1 top). The magnetic structure is collinear FM with the Ce magnetic moments strongly reduced (0.6 $\mu_{B}$) and lying along the b-axis. Both $\chi^{\prime}_{ac}$ and $\chi^{\prime\prime}_{ac}$ curves (displayed in Figure 1 top) show a pronounced maximum at $T_{f}$, which shifts with the frequency of the ac applied field, and is associated to a cluster-glass state \cite{Binder,Mydosh}. The specific heat, also plotted in the same figure, exhibits a broad anomaly near $T_{f}$. Such cluster-glass state is further supported by the irreversibility of the field cooled and zero field cooled magnetic susceptibility (see Figure 1 bottom), although it starts at $T^{*} > T_{f}$. This last observation implies the presence of short range correlations well above $T_{f}$. 

The $\mu$SR experiments have been essential in order to interpret these results. The characteristic parameters obtained from such a study are summarized in the inset of Figure 1 bottom. The most significant result is the existence of an intermediate state developing at $T_{f}<T<T^{*}$ and consisting of long-range ordered and non-ordered phases, the former increasing as the temperature is gradually decreased. This situation, described as the Griffiths phase \cite{CastroNeto} was also detected by $\mu$SR in the NFL CeCoGe$_{1.8}$Si$_{1.2}$ compound \cite{Krishnamurthy}. Below $T_{f}$, muons detect long-range order with a broad molecular field distribution on the muon site, which indicates the presence of strong local spin disorder in this state. It has to be stressed that no magnetic contribution (within the experimental resolution limits) is observed in the neutron diffraction pattern obtained just below $T_{f}$. In fact, $\mu$SR is particularly sensitive to short-range order correlations \cite{Kalvius} and therefor, a shorter coherence length (as compared to neutron diffraction) suffices  to define the long-range ordered state from  $\mu$SR. The overview of Figure 1 also reveals the absence of a transition temperature below $T_{f}$. Thus, the main question arising from this puzzling phenomenology is the mechanism that leads the system from a cluster-glass into a long-range ordered state. In order to answer this crucial question, and bearing in mind all the previously described results, we propose a phenomenological picture that takes into account all the experimental evidence.

Such a picture is shown in Figure 2 and illustrates the magnetic state of the system in different temperature regimes. As the temperature is lowered from the paramagnetic state, regions where the magnetic moments fluctuate together, or clusters, develop due to the rise in short-range magnetic interactions. The temperature that corresponds to the cluster-formation ($T^{*}$) corresponds to the upper boundary of the intermediate inhomogeneous state detected by $\mu$SR. The volume fraction of these dynamic entities increases when the temperature decreases and they freeze at $T_{f}$, as seen by $\chi_{ac}$, $C_{P}$ and dc-magnetization. At this temperature, and just below, magnetic correlations are large enough to be considered as long-range order by $\mu$SR but not by neutron diffraction. Very low temperatures are required ($T \ll T_{f}$) in order to detect magnetic contribution in the neutron diffraction pattern. Taking into account this body of evidence, we propose a percolative process in order to describe the emergence of the long-range ferromagnetic state from the cluster-glass one below $T_{f}$. According to such a mechanism, the size of the magnetic clusters would increase below $T_{f}$, leading to a domain-like ferromagnetic state at very low temperature.

The proposed percolative scenario is strongly supported by two recent experiments. Their results are described below.

\begin{figure}
\includegraphics[width=7cm]{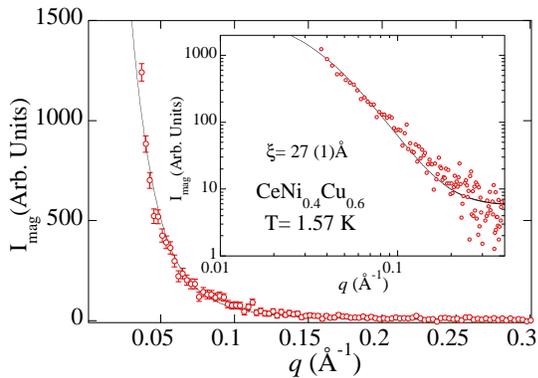}
\caption{Magnetic SANS intensity as a function of the neutron momentum transfer, $q$, for CeNi$_{0.4}$Cu$_{0.6}$ compound at 1.57 K, just below $T_{f}$. The solid lines are a fit to a Lorentzian-squared $q$ dependence (see text). $\xi$ corresponds to the estimated correlation length obtained from the fit. The inset shows the same fit in a log-log scale.}
\end{figure}

The most direct evidence one can gain of the existence of magnetic clusters is obtained by SANS. For this reason, we have carried out SANS measurements for several samples of the series using the D16 instrument located at the Institute Laue-Langevin (ILL) in Grenoble. The SANS magnetic signal was obtained at temperatures below the $T_{f}$ detected by macroscopic measurements, and the incoherent nuclear scattering was removed by subtracting the SANS signal recorded in the paramagnetic regime ($T > T^{*}$). The results obtained for the x= 0.6 sample are presented in Figure 3 as an example. When analyzing the small angle magnetic scattering cross section of a sample exhibiting inhomogeneous states, such as cluster-glass behaviour, one would expect to have two different contributions: a Lorentzian term [$I=A/(\emph{q}^{2}+\emph{1/$\xi$}^{2})$] representing the short-range ferromagnetic order associated to the fluctuations of the spin system and characteristic of temperatures around the Curie temperature, and a Lorentzian-squared one [$I=B/(\emph{q}^{2}+\emph{1/$\xi$}^{2})^{2}$] arising from scattering from static regions of local spin ordering \cite{Birgeneau,Hellman}. However, in the present case, only the latter is needed to account for the magnetic SANS signal. The conclusions that can be obtained from these results are: (i) The direct observation of clusters of magnetic origin and the clear evidence that long-range magnetic order still has not been fully established at the lowest measured SANS temperature (1.57  K for x = 0.6); (ii) We have obtained a correlation length of 27 \AA , which is related with the average size of the magnetic clusters. It is worth noting that the results obtained for the x= 0.2 sample  give much smaller values of the magnetic SANS intensity and correlation length, as is expected due to the highly reduced magnetic moment in this composition as a consequence of the enhancement of the Kondo effect and weaker magnetic interactions.

\begin{figure}
\includegraphics[width=7cm]{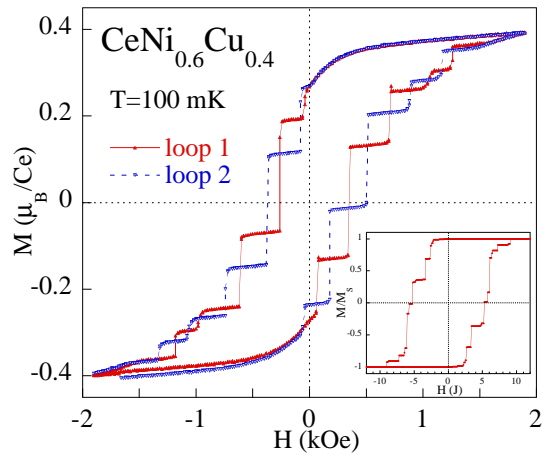}
\caption{Hysteresis loops obtained for CeNi$_{0.6}$Cu$_{0.4}$ at 100 mK.  The inset shows the numerical simulation of a hysteresis loop (see details in the text). The magnetization is in relative units and the magnetic field in units of the exchange constant, $J$.}
\end{figure}

Magnetization was measured down to 100 mK ($T \ll T_{f}$) using a SQUID magnetometer equipped with a miniature dilution refrigerator at the CRTBT (CNRS, Grenoble). Two representative compounds with x = 0.4 and 0.5 were studied, both of which display long-range ferromagnetic order at this low temperature as detected by neutron diffraction measurements. Figure 4 shows two different hysteresis loops obtained for CeNi$_{0.6}$Cu$_{0.4}$ at T = 100 mK. The magnetization changes in a series of large discrete jumps giving rise to a multistep pattern. When repeating the hysteresis loop measurement, the steps appear at different field values. This unexpected staircase-like behaviour only appears in the very low temperature loops (well below $T_{f}$), vanishing for slightly higher temperatures (300 mK). Similar results have been obtained in CeNi$_{0.5}$Cu$_{0.5}$. The observed features in these cycles, clearly related to avalanches of domain flips, are the mesoscopic analogue of the Barkhausen noise \cite{Lhotel,Tung}. 

The magnetic domains appear in conventional ferromagnets in order to minimize the magnetostatic energy. In the present case, they are a consequence of the thermally activated percolative process of static ferromagnetic clusters reaching a minimum energy state. This mechanism increases the magnetic correlation length up to values that can be detected by neutron diffraction ($\sim 10^3$ \AA ). The process is driven by the increasing importance of the RKKY interaction as the temperature decreases. This interaction, then, competes with the local anisotropy, giving rise to a structure of magnetic domains that displays an "asperomagnetic" mesoscopic state such has been reported by Coey \cite{Coey} in the case of amorphous systems. The present situation is clearly reminiscent of that case, but occurring in crystalline samples. 

Recent simulations based on disorder, anisotropy and competing magnetic interactions reproduce satisfactorily the experimental situation. We have used an Ising-like Hamiltonian with a positive exchange interaction and an anisotropy term that has been fixed at random for each one of the random size clusters in which the lattice has been split. Furthermore, in order to simulate the disorder in the interactions, we have ``isolated'' some spins (typically 2.5\%), by eliminating the ferromagnetic link with their neighbours. Within this model, we have performed a Monte Carlo simulation on a 3-dimensional lattice at zero temperature. One typical hysteresis cycle so obtained is shown in the inset of Figure 4, exhibiting remarkable resemblance with the experimental results.

In conclusion, the use of many different techniques has allowed us to propose a mesoscopic cluster-glass state and to extend the percolative scenario, already described in manganites \cite{Burgy03}, magnetic semiconductors \cite{Mayr} and diluted magnets \cite{Vojta}, to strongly correlated electron metallic systems and, in particular, to Ce and U based intermetallic compounds. This model has been confirmed by SANS and the staircase hysteresis loops at low temperatures and represents, in our opinion, a more general situation than the particular case of CeNi$_{1-x}$Cu$_{x}$ and certainly shed light on the nature of magnetic ground state of other compounds with small magnetic moments, weak ferromagnetism and intrinsic disorder effects \cite{Young,Bauer,Coleman}.

This work is supported by the MAT2003-06815 project and the ECOM COST Action P16. N. Marcano acknowledges the Spanish MEC for financial support. We thank G.M. Kalvius, L. Fernandez Barquin, J. Rodriguez Fernandez, B. Coqblin, S. Magalhaes, V. Sechovsky, M.B. Maple, G.G. Lonzarich, P. Haen and S.S. Saxena for useful discussions. ILL staff is acknowledged for support with the SANS measurements.

\end{document}